# Superconducting nanowire single-photon detectors: physics and applications

**Chandra M Natarajan[1, 2], Michael G Tanner[1] and Robert H Hadfield[1]**

[1]Scottish Universities Physics Alliance and School of Engineering and Physical Sciences, Heriot-Watt University, Edinburgh, EH14 4AS, UK

[2]E. L. Ginzton Laboratory, Stanford University, Stanford, California 94305, USA

Email: r.h.hadfield@hw.ac.uk

**Abstract:** Single-photon detectors based on superconducting nanowires (SSPDs or SNSPDs) have rapidly emerged as a highly promising photon counting technology for infrared wavelengths. These devices offer high efficiency, low dark counts and excellent timing resolution. In this Review, we consider the basic SNSPD operating principle and models of device behaviour. We give an overview of the evolution of SNSPD device design and the improvements in performance which have been achieved. We also evaluate device limitations and noise mechanisms. We survey practical refrigeration technologies and optical coupling schemes for SNSPDs. Finally we summarise promising application areas, ranging from quantum cryptography to remote sensing. Our goal is to capture a detailed snapshot of an emerging superconducting detector technology on the threshold of maturity.

**Contents**

# 1   Introduction

*1.1   Detection of optical radiation with superconductors*

The remarkable phenomenon of superconductivity was discovered a century ago by Heike Kammerlingh Onnes [1]. This breakthrough opened the pathway to a tantalizing range of real-world applications. The superconducting state is sensitive to incident radiation at optical wavelengths [2], and the advent of thin film superconductors, microfabrication techniques and laser sources allowed the first superconducting radiation detectors and bolometers to be demonstrated as referenced in [3]. Spurred by the demands of fields such as astronomy, a suite of single-photon sensitive energy resolving superconducting detectors have been developed operating at sub-Kelvin temperatures: the superconducting tunnel junction (STJ) [4], the transition edge sensor (TES) [5] and the kinetic inductance detector (KID) [6].

Just a decade ago a new superconducting device concept was demonstrated by Gregory Gol'tsman and colleagues, based on a niobium nitride nanowire [7]. This type of device, known as the superconducting single-photon detector (SSPD) or superconducting nanowire single-photon detector (SNSPD), is single-photon sensitive at visible and infrared wavelengths, with recovery times and timing precision orders of magnitude faster than existing single-photon detectors based on superconducting materials. In addition, this detector operates at the boiling point of liquid helium (4.2 K), a temperature now within reach of rapidly improving closed-cycle cooling technology [8]. SNSPDs have excellent potential for time-correlated single-photon counting (TCSPC) [9] in the infrared wavelength regime, where important new applications are emerging. The main competitors in this arena are solid-state single photon avalanche photodiodes (SPADs), which have superseded bulky photomultipliers (PMTs) [10]. The long wavelength sensitivity of the SNSPD extends far beyond that of the Si single photon avalanche photodiode (SPAD) [11] and the SNSPD is superior to the InGaAs SPAD [12] in terms of signal-to-noise ratio. SNSPDs have been the subject of intense interest of over the past decade and many research groups around the world have contributed to device development. This review aims to summarize the basic device operating principle, evolution of SNSPD design, refrigeration and materials considerations and to give an overview of promising applications.

*1.2   Single-photon detection: basic principles and metrics*

An ideal single-photon detector (SPD) generates an electrical signal only upon absorption of a photon. The signal level is well defined above the noise; in the absence of illumination no electrical signal is returned. In practice, real SPDs have many non-ideal characteristics and performance metrics must be carefully defined. The most obvious performance metric in a SPD is the detection efficiency ($\eta$) – the probability that an output signal is registered when a photon is incident on the detector. In practice $\eta$ is lower than 100% and is likely to depend strongly on the wavelength $\lambda$ of the incident photons. In any real photon-counting experiment, photons can be lost before reaching the detector due to absorption, scattering or reflection within the experimental environment – this loss can be defined as the *coupling efficiency* ($\eta_{coupling}$). The detector material and geometry determines the *absorption efficiency* ($\eta_{absorption}$). Finally, there may be a non-unity probability that the detector generates an output electrical signal after photon absorption – we define this as the *registering probability* ($\eta_{registering}$). Taking all these contributions into account, the overall *system detection efficiency* ($\eta_{sde}$) is

$$\eta_{\text{sde}} = \eta_{\text{coupling}} \times \eta_{\text{absorption}} \times \eta_{\text{registering}} \quad (1)$$

The intrinsic device detection efficiency ($\eta_{\text{dde}}$) is defined as

$$\eta_{\text{dde}} = \eta_{\text{absorption}} \times \eta_{\text{registering}} \quad (2)$$

Therefore $\eta_{\text{sde}}$ and $\eta_{\text{dde}}$ are only equal for perfect optical coupling i.e. $\eta_{\text{coupling}} = 1$. The device detection efficiency, $\eta_{\text{dde}}$, is often reported as 'quantum efficiency' in the literature. However this term is used with different meanings for different technologies, so 'device detection efficiency' will be used in this text for clarity. Readers should also heed the following caveat: unfortunately across the published SPD literature there is no uniform consensus on whether to report $\eta_{\text{sde}}$ or $\eta_{\text{dde}}$. In general early SPD device demonstrations report $\eta_{\text{dde}}$, whereas as when SPDs are implemented in actual photon-counting experiments, $\eta_{\text{sde}}$ is more likely to be reported.

Other performance metrics quantify limitations in SPD performance. Stray light and electrical noise can also potentially mimic the optical signal. These false detection events are called *dark counts,* usually quantified in terms of a dark count rate (DCR). The timing uncertainty between the arrival of the photon at the SPD (which can be known extremely precisely with modern optical sources) and the generation of the output pulse from the SPD, may set the timing resolution. This is known as the *jitter* ($\Delta t$) of the SPD. In addition, a real SPD will have a finite *recovery time* ($\tau$) before it is ready to register a subsequent photon – this sets a limit on the theoretical maximum count rate of the SPD (which in practice may be limited by other factors, such as the readout electronics).

## 2   Superconducting Nanowire Single-Photon Detectors (SNSPDs)

### 2.1   *Origins of device concept*

In 1971, a laser was used for the first time to disrupt the superconductivity of Pb films, forming a resistive state that could not be explained just by heating effects [2]. This experiment showed that the energy of the absorbed photons resulted in a nonequilibrium state, with hot excited quasiparticles at a higher temperature than the Cooper pairs in the superconductor. Equilibrium is reached via a sequence of relaxation processes: (i) inelastic scattering of quasiparticles by electron-electron interactions and electron-phonon interactions, (ii) generation of quasiparticles by phonons, (iii) quasiparticle recombination and (iv) energy dissipation in to the substrate by the phonon in the superconductor. The temporal characteristics of the relaxation depend on the time constants associated with each of these processes. Twenty years later, an elegant approach to understand the nonequilibrium relaxation was developed by studying transient photoimpedence response (TPR) of superconducting films [13]. In a TPR measurement the superconducting film is approximated as a changing kinetic inductance and the film in the resistive state is approximated as a photoresistor. A two-temperature model is used to quantitatively extract the time constants from the photoresponse of a superconductor film to an optical pulse [14]. In this model, the time-dependent temperature of the electron subsystem ($T_e$) and phonon subsystem ($T_p$) are obtained by solving the following coupled linear heat-balance equations:

$$c_e \frac{dT_e}{dt} = -\frac{c_e}{\tau_{e\text{-}p}}\left(T_e - T_p\right) + P(t),$$

$$c_p \frac{dT_p}{dt} = -\frac{c_e}{\tau_{e\text{-}p}}\left(T_e - T_p\right) - \frac{c_p}{\tau_{es}}\left(T_p - T_0\right)$$

(3)

where $c_e$ and $c_p$ are the electron and phonon specific heat, respectively, $T_0$ is the substrate temperature, $\tau_{e\text{-}p}$ is the electron-phonon interaction time, $\tau_{es}$ is the phonon escape time and $P(t)$ is the time-dependent power of radiation absorbed in the unit volume of the film. This two-temperature model ignores any Joule heating via the bias current. This hot-electron and phonon-heatsink scheme predicts a picosecond photoresponse from the superconductor [14]. This was experimentally verified in NbN [15] and YBa$_2$Cu$_3$O$_{7\text{-}\delta}$ [16] microbridges. The region of hot electrons or quasiparticles forming the resistive region in the superconductor is known as the 'hotspot'. The hotspot concept was originally developed by Skocpol *et al* to study self-heating effects in superconducting microbridges [17]. In 1996, Kadin *et al* [18] approximated the two-temperature model equations as a heat flow equation:

$$Cd\frac{\partial T}{\partial t} = \kappa d \nabla^2 T + \alpha(T_0 - T)$$

(4)

where $d$ is the thickness of the nanowire, $\kappa$ is the thermal conductivity, $C$ [$C = c_e + c_p$] is the heat capacity of the superconducting film, $\alpha$ is the thermal boundary conductance between the film and the substrate, $T_0$ is the substrate temperature (since the nanowire is thin enough) and the temperature of the nanowire is approximated to one temperature $T$. The solutions to equation 4 predict that the photon-induced hotspot is on the few nanometre to tens of nanometre scale. Therefore it was proposed that patterning the superconducting film into a device with comparable dimensions to the size of a hotspot would result in an infrared sensitive SPD [18].

In 2001, Gregory Gol'tsman and coworkers [7] demonstrated the single-photon sensitivity ($\lambda$=790 nm) of a current-biased NbN superconducting nanowire (200 nm wide, 5 nm thick, 1 µm long) maintained at 4.2 K. This device is known as the superconducting single-photon detector (SSPD) or superconducting nanowire single-photon detector (SNSPD).

## 2.2  SNSPD device physics

### 2.2.1  Device operating principle

The basic device operation of the Superconducting Nanowire Single-Photon Detector (SSPD/SNSPD) can be described as follows [7, 19, 20]: the NbN nanowire is maintained well below its superconducting critical temperature $T_c$ and direct current biased just below its critical current (Figure 1(a)(i)). Individual infrared photons have enough energy to disrupt hundreds of Cooper pairs in a superconductor thereby forming a hotspot (Figure 1(a)(ii)). The hotspot itself is not large enough to span the width of the ~100 nm nanowire. The hotspot region forces the supercurrent to flow around the resistive region (Figure 1(a)(iii)). The local current density in the sidewalks increases beyond the critical current density and forms a resistive barrier across the width of the nanowire. The sudden increase in resistance from zero to a finite value generates a measurable output voltage pulse across the nanowire (Figure 1(a)(iv)).

*2.2.2 A phenomenological model of SNSPD operation*

A simple but elegant phenomenological model can be used to simulate the measurable output voltage pulse from an SNSPD biased at $I_{bias}$ as shown in Figure 1(b) [21, 22]. An inductor $L_k$ and a resistor $R_n(t)$ represent the kinetic inductance of the superconducting nanowire and the time dependent hotspot resistance respectively. The kinetic inductance is given as $L_k = \mathcal{L}_k \int ds/A(s)$ where $\mathcal{L}_k$ is the kinetic inductivity and $A$ is the cross-sectional area. Absorption of a photon forms a resistive barrier, which can be simulated as the instantaneous opening of the switch. Introducing $R_n(t)$ into the circuit diverts the current into the load impedance $Z_0$ (typically $50\Omega \ll R_n(t)$). The LR circuit imposes a time constant of $\tau_1 = L_k/(Z_0 + R_n(t))$ on the decay of $I_{bias}$ through the nanowire. This $\tau_1$ limits the rise time of the voltage pulse across $Z_0$. The simulated voltage pulse is shown in Figure 1(c). As the current flowing through the nanowire decays to a return current ($I_r$), the electrothermal feedback reduces the Joule heating of the resistive region by the bias current in the nanowire. In this model, the switch closes again to remove $R_n(t)$ from the circuit, representing the initiation of the recovery of supercurrent in the nanowire. The current recovers from $I_r$ to $I_{bias}$ with a longer time constant $\tau_2 = L_k/Z_0$ due to the reduction of total resistance in the circuit (dashed red curve in Figure 1(c)). The SNSPD remains insensitive to photons while the supercurrent decays and eventually recovers to a re-triggerable level. The overall non-sensitive time or dead time of the SNSPD is given by $\tau = \tau_1 + \tau_2 \approx \tau_2$. $\tau$ can be reduced considerably by either reducing the value of $L_k$ (either by the use of a shorter or thicker nanowire) [22] or including a resistance $R_s$ in series to the device, thereby reducing the decay time [20]. The kinetic inductance ($L_k$) of the nanowire is directly related to its length ($l$), cross-sectional area ($A$) and magnetic penetration depth ($\lambda$) from simplified one-dimensional Ginzberg-Landau theory [$L_k = (\mu_0\lambda^2)(l/A)$]. The magnetic penetration depth ($\lambda$) depends on the critical temperature ($T_c$), the normal state resistivity ($\rho_{T>T_c}$) of the nanowire and the operating temperature – therefore the choice of material (see section 2.3.5) and the operating temperature will influence $L_k$.

*2.2.3 Refinements to the electro-thermal model*

Recently, closer attention has been paid to whether the hotspot resistance is fixed or evolves with time. The measured hotspot resistance is considerably higher than the value predicted by this simple hotspot theory. Growth of the resistive region due to Joule heating accounts for the high resistance of the nanowire (Figure 1(a)((v)) [20]. Equation 4 can be modified to include Joule heating as follows:

$$Cd\frac{\partial T}{\partial t} = J^2 d\rho + \kappa d\nabla^2 T + \alpha(T_0 - T) \qquad (5)$$

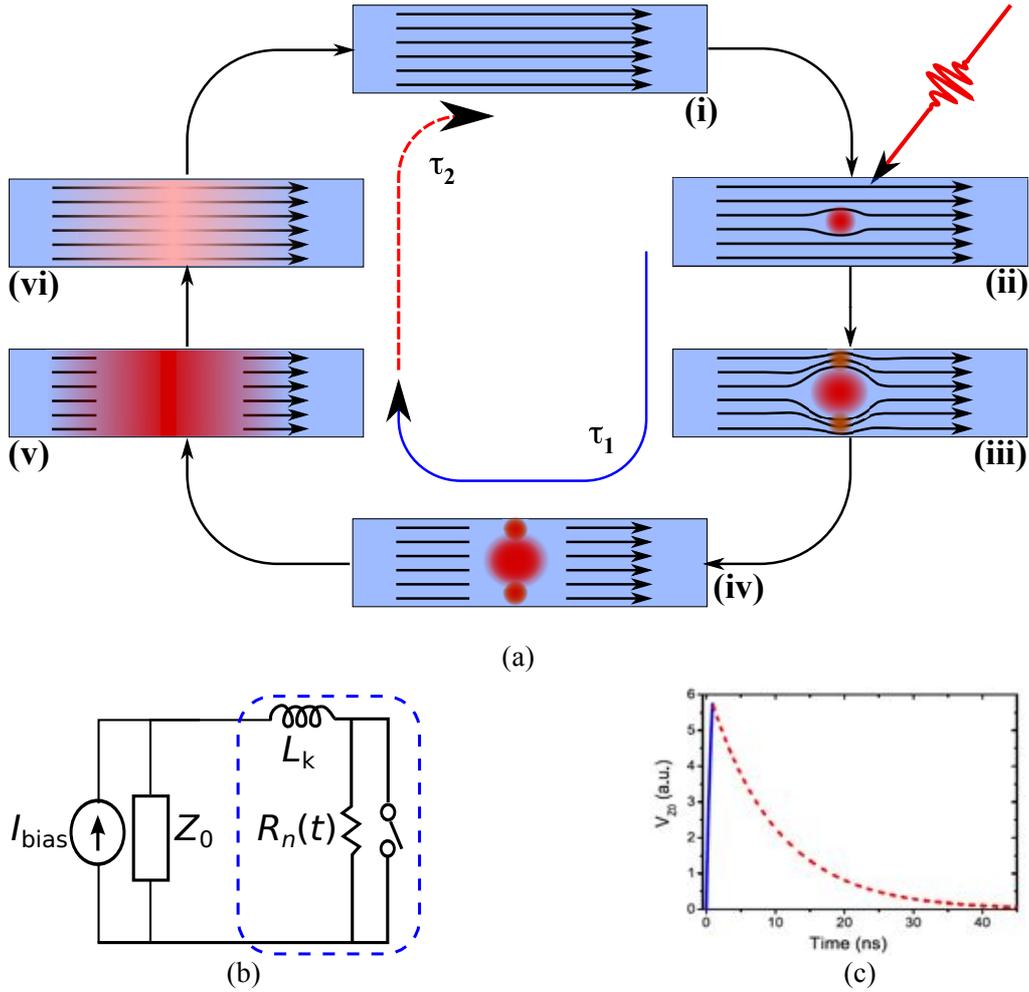

**Figure 1.** The basic operation principle of the superconducting nanowire single-photon detector (SNSPD) (after Gol'tsman [7], Semenov et al [19] and Yang [20]): (a) A schematic illustrating the detection cycle. (i) The superconducting nanowire maintained well below the critical temperature is direct current (DC) biased just below the critical current. (ii) When a photon is absorbed by the nanowire creating a small resistive hotspot. (iii) The supercurrent is forced to flow along the periphery of the hotspot. Since the NbN nanowires are narrow, the local current density around the hotspot increases, exceeding the superconducting critical current density. (iv) This in turn leads to the formation of a resistive barrier across the width of the nanowire [7]. (v) Joule heating (via the DC bias) aids the growth of resistive region along the axis of the nanowire [20] until the current flow is blocked and the bias current is shunted by the external circuit. (vi) This allows the resistive region to subside and the wire becomes fully superconducting again. The bias current through the nanowire returns to the original value (i). (b) A simple electrical equivalent circuit of a SNSPD. $L_k$ is the kinetic inductance of the superconducting nanowire and $R_n$ is the hotspot resistance of the SNSPD. The SNSPD is current biased at $I_{bias}$. Opening and closing the switch simulates the absorption of a photon. An output pulse is measured across the load resistor $Z_0$ [21, 22]. (c) A simulation of the output voltage pulse of the SNSPD (approximating the pulse shape typically observed on an oscilloscope after amplification). Values of $L_k$ = 500 nH and $R_n$ = 500 Ω have been used for this simulation (for simplicity the $R_n$ is assumed fixed, although a more detailed treatment [20] shows $R_n(t)$). The solid blue line is the leading edge of the SNSPD output pulse, whilst the dotted red line is the trailing edge of the output pulse. The time constants relate to the phases of the detection cycle in (a).

where $J$ is the current density through the wire and $\rho$ is the electrical resistivity. The hotspot cools down by coupling the energy of the excited electrons to the phonons via electron-phonon scattering, with a time constant $\tau_{e-p}$ (~ 10 ps). The phonon-phonon scattering then couples the energy into the substrate with a time constant $\tau_{p-sub}$. A fraction of the energy is backscattered into the electron system owing to mechanisms such as the lattice mismatch between the superconducting nanowire and the substrate. The substrate acts as a heat sink at temperature $T_0$ [23]. Then the nanowire recovers its superconducting state (figure 1(a)(vi)). The energy of the incident photon is negligible compared to the energy stored in the kinetic inductance; therefore energy or photon-number resolution is lost due to the Joule heating of the hotspot [19, 20].

*2.2.4 Modelling device performance metrics*

The phenomenological model described above gives insight into some of the SNSPD performance metrics, namely the device recovery time ($\tau$), timing jitter ($\Delta t$) and dark count rates, discussed further in sections 2.3.4 and 2.3.8. The other paramount performance metric is the system detection efficiency $\eta_{sde}$. Equation (1) specifies the factors that must be considered, however the model described only affects the third term, $\eta_{registering}$, the probability that a pulse is emitted from the SNSPD (assuming the associated counting electronics in the system register every pulse from the detector). Improvements in the registering probability are discussed in section 2.3.3.

For the SNSPD, $\eta_{coupling}$ describes the challenge of efficiently illuminating the potentially sub-micrometer device area (discussed in section 2.3.1 and 2.4.2). If the illumination is perpendicular to the device surface $\eta_{absorption}$ describes the probability that the photon is absorbed into the superconducting nanowire as opposed to being transmitted through or reflected by the thin film and any surrounding optical structure. This can be simply modelled as a solution of Fresnel equations for light moving between media of differing refractive indices if the optical properties of the substrate and superconductor are known [24]. Progress in increasing the absorption probability is discussed in section 2.3.2. The sub 100 nm fine structure of the detector (commonly an order of magnitude less than the wavelength of the incident light) is a small perturbation of this result [25], described in section 2.3.7.

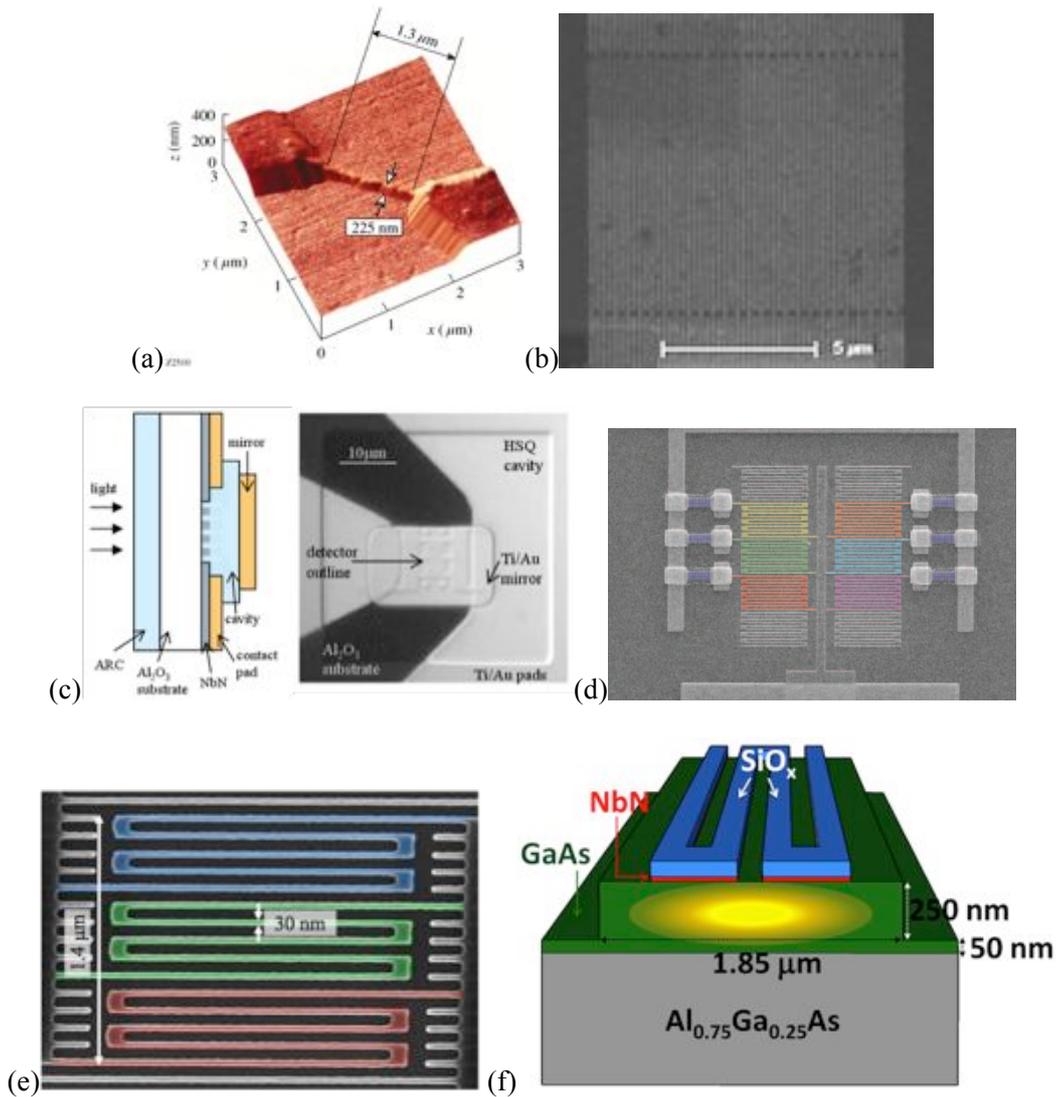

**Figure 2.** The evolution of SNSPD design (a) An atomic force microscopy (AFM) image of an early 1.3 μm x 225 nm NbN nanowire (after [26]), (b) A scanning electron micrograph (SEM) of NbN meander SNSPD covering a 10 μm x 10 μm area (of the type first presented in [27]). (c) A 3 μm x 3 μm NbN meander SNSPD embedded in an optical cavity, designed for optical illumination via the substrate [28]. (d) Multiple nanowire elements are biased in parallel via independent resistors resulting in a photon-number-resolving SNSPD (PNR-SNSPD) [29]. (e) Ultrathin nanowires (30 nm wide) are connected in parallel to improve the sensitivity (registering efficiency) of the SNSPD – this device is known as a superconducting nanowire avalanche photodetector (SNAP) [30] (f) A SNSPD fabricated on an optical waveguide structure to improve the optical coupling efficiency (after [31]). Figure (a) reproduced with permission from IEEE [26]. (b) reproduced with permission from Nature Publishing Group [102]. (c) reproduced with permission from The Optical Society of America [28]. (d) reproduced with permission from Nature Publishing Group [29]. (e) reprinted with permission from Marsili F et al 2011 Nano Lett. **11** 2048. Copyright 2011 American Chemical Society [30]. (f) reprinted with permission from Sprengers J P et al 2011 Appl. Phys. Lett. **99** 181110. Copyright 2011, American Institute of Physics [31].

## 2.3 Evolution of SNSPD Devices

Since the demonstration of the first SNSPD a decade ago (Figure 2(a)), many research groups around the world have taken up the challenge of improving SNSPD operation. The goal has been to improve the performance metrics described in section 1.2. A variety of strategies have been employed (Figure 2). Large area 'meander SNSPDs' (Figure 2(b)) have been developed to improve the coupling efficiency and hence the practical detection efficiency. Cavity and waveguide integrated designs (Figures 2(c) and 2(f) respectively) have been employed to boost the absorption efficiency. Multipixel device designs (Figure 2(d)) have been used to achieve photon number resolution. Ultranarrow wires (Figure 2(e)) and alternate superconducting materials have been employed to improve the registering efficiency. As new device designs have been evaluated, important limiting factors such as constrictions and latching have been identified and understood.

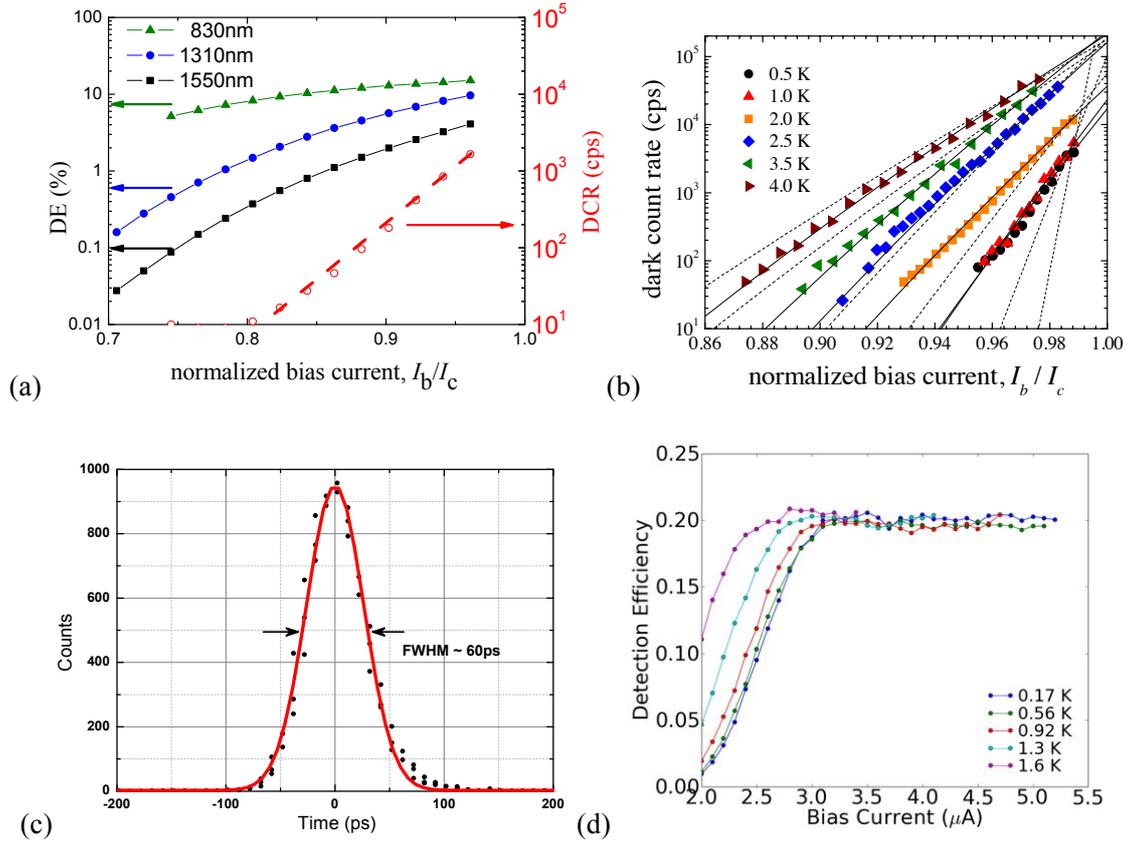

**Figure 3.** The characteristics of SNSPDs (a) the system detection efficiency ($\eta_{sde}$) of a fibre-coupled 20 μm x 20 μm area NbN SNSPD at ~3 K for λ=830 nm, 1310 nm and 1550 nm. This device is of the type reported in [32, 33], (b) The variation in dark count rate (DCR) with temperature [34], (c) The instrument response function of a NbTiN SNSPD measured using a femtosecond laser and TCSPC card at λ=1550 nm. The $\Delta t$ of 60 ps FWHM (determined via a fitted Gaussian function) is quoted as the jitter of the device in [35], (d) A SNSPD fabricated in a low energy gap superconductor, a-$W_xSi_{1-x}$. A clear saturation of $\eta_{sde}$ (λ=1550 nm) occurs at higher bias currents [36]. Figure (b) reprinted with permission from Yamashita T et al 2011 Appl. Phys. Lett. **99** 161105. Copyright 2011, American Institute of Physics [34]. Figure (d) reproduced with permission from Baek B et al 2011 Appl. Phys. Lett. **98** 251105. Copyright 2011, American Institute of Physics [36].

*2.3.1 Boosting the coupling efficiency*
Focusing an optical spot at λ = 1550 nm on to a 150-200 nm wide nanowire [26] (Figure 2(a)) with $\eta_{coupling}$ approaching unity is not feasible. In order to enhance $\eta_{coupling}$, a nanowire meander is written typically across a 10 μm x 10 μm [27, 37] (figure 2(b)) or 20 μm x 20 μm area [32]. These NbN meander devices show system detection efficiencies ($\eta_{sde}$) of about 2-3% at 1550 nm, 1 kHz DCR at T~3 K (Figure 3(a)). As illustrated in Figure 3(a), in general for SNSPDs $\eta_{sde}$ decreases as the energy of the photon decreases (i.e. at longer wavelength) – pointing to a decrease in $\eta_{registering}$ [27, 38]. Figure 3(c) shows typical timing jitter (60 ps FWHM) for a large area SNSPD measured with TCSPC electronics and a femtosecond laser.

### 2.3.2 Enhancing absorption efficiency

Intrinsic device detection efficiencies ($\eta_{dde}$) of 57% for 1550 nm photons (and 67% at 1064 nm), at 1.8 K, have been demonstrated by embedding a 3 μm x 3 μm nanowire meander SNSPD inside an optical cavity in order to increase $\eta_{absorption}$ (Figure 2(c)) [28]. This design has been refined further (via a buried oxide and a four pixel layout) to achieve 87% $\eta_{dde}$ at 1550 nm [39]. Efficient optical coupling is required to convert these intrinsic values into practical system efficiencies, as discussed in section 2.4.2. Impressive system efficiency results have been reported with $\eta_{sde}$ ~ 24-28% [40, 41], ~ 23-40% [35, 40] and ~ 46% [42] for 1550, 1310 and 900 nm photons respectively. Recently system detection efficiency as high as $\eta_{sde}$ = 73% for a 14 μm diameter 4 element detector operating for 1550 nm photons has been presented [43].

Optical coupling via lensed fibre [40, 41] or substrate thinning [44] is required to focus the light to the device active area to minimise the beam divergence in the substrate if a back side coupling regime is implemented. On the other hand, superconducting nanowires can be patterned on top of wavelength specific $SiO_2$/Si [35, 45] or GaAs/AlGaAs distributed Bragg mirrors [46]. The advantage here is that front side fibre coupling is sufficient to enhance absorption for these devices. The coupling can be further improved by integrating an Au optical antenna with the nanowires [47].

Recently several groups have proposed [48, 49] or demonstrated [31, 50-52] embedding the nanowire in an optical waveguide [31, 48-52] (an example is illustrated in Figure 2(f)). A nanowire running along the length of an optical waveguide [31, 51] should give a long interaction length for incident photons, allowing $\eta_{absorption}$ to be maximised. An array of nanowires can be used to sample a standing wave in an optical waveguide interferometer for spectroscopy applications [50, 52].

### 2.3.3 Improving registering efficiency

$\eta_{registering}$ of the device can be improved by reducing the width of the nanowires. Multiple narrow wires can be placed in parallel to boost the output signal whilst improving long wavelength sensitivity [53]. Ultra-narrow nanowire (20 nm or 30 nm) devices (Figure 2(e)) have been demonstrated to be more responsive to low-energy photons than typical nanowire devices (90 nm) [30]. In this work four ultra-narrow nanowire elements are placed in parallel, and the device is referred to as a Superconducting Nanowire Avalanche Photodetector (SNAP). Maximum $\eta_{registering}$ is observed as $\eta_{dde}$ saturates at a lower bias current. Upon absorption of a photon in a single parallel element of the device, the element turns resistive and diverts additional current into the other elements, switching the entire device resistive by exceeding the critical current of the parallel elements [54]. If the bias current is significantly lower than the critical current, the first photon absorption switches the first element but the diverted current may not be enough to trigger the avalanche. In this scenario, the device waits either for a second photon to be absorbed in another element or the occurrence of a dark count in another element to switch the whole device resistive. If this does not occur the device can recover without completing the avalanche and registering a count [54]. The registering probability saturation is also observed when smaller superconducting gap energy materials such as a-$W_xSi_{1-x}$ [36] (figure 3(d)) are used for fabrication of SNSPDs. These devices exhibit $\eta_{dde}$ saturation (and near unity $\eta_{registering}$) at λ=1550 nm even when the nanowires are 150 nm wide with 4.5 nm film thickness. This behaviour is most likely a consequence of the smaller size of superconducting energy gap in this material leading to an increased hotspot size for a given photon energy.

### 2.3.4 Constrictions, latching and recovery time

Fabricating uniform, large area meanders [27, 32, 37, 55] or meanders with fill factors > 50% [56] with high yield remains a challenge. Constrictions in the nanowire do not allow the detector to be biased at a high current, consequently lowering $\eta_{dde}$ of the device due to a lower $\eta_{registering}$ [57]. Nano-optical studies have been undertaken to confirm that a constricted device is sensitive only at the constriction and is insensitive or less sensitive at the non-constricted regions due to lower current density [58]. For a given defect density, reducing the area of nanowire meander will improve the nanowire uniformity. 3 μm x 3 μm devices have been shown to be highly uniform exhibiting higher $\eta_{dde}$ [57]. However, coupling the light efficiently to this small 3 μm x 3 μm area is challenging. $L_k$ is lower for shorter nanowires, resulting in a shorter $\tau$. This makes shorter nanowires highly suitable for high speed applications. The load impedance can also be increased to reduce $\tau$. However, $\tau$ suppression has a limitation. If $\tau$ is reduced excessively, the current will return to the nanowire too rapidly. Joule heating results in a self heating hotspot [59]. This self-heating hotspot causes the nanowire to enter into a "latched" state, preventing further detection of photons without manually reducing $I_{bias}$. Alternatively, the inductance of the detector can be reduced by writing a set of parallel nanowires [60-62] or an array of parallel nanowire pixels [63] rather than relying on a single meander nanowire across a given area.

### 2.3.5 Alternate superconducting materials for SNSPDs

NbN was the first choice material for SNSPDs. Subsequently SNSPDs have also been demonstrated using NbTiN (a superconducting material with very similar properties to NbN) on silicon based substrates [64, 65]. NbTiN nanowire devices have been shown to possess low $L_k$ due to their low resistivity above the superconducting transition temperature, therefore NbTiN SNSPDs have smaller $\tau$ compared to the NbN devices [65]. Similarly Nb SNSPDs have lower resistivity than NbN resulting in shorter $\tau$, but the Nb devices suffer from latching due to their slow energy relaxation process [66]. Fabricating a superconducting nanowire meander using novel or high-$T_c$ superconductors is arguably just a question of technological development. Recently $MgB_2$ SNSPDs have been fabricated [67] and single-photon sensitivity at visible wavelengths has been demonstrated [68], but new fabrication and film growth techniques have to be developed to increase the active area by patterning uniform meanders without defects. Ultra-thin nanowires have been processed in $YBa_2Cu_3O_{7-\delta}$ thin films but single-photon sensitivity is yet to be demonstrated at visible or infrared wavelengths [69]. Fabrication challenges aside, this result may be expected as high-$T_c$ superconducting materials have a large superconducting gap energy compared to NbN, reducing the sensitivity to photons of a given energy. Single-photon sensitivity has been reported in NbN SNSPDs up to 5 μm wavelength, albeit with a significant reduction in $\eta_{dde}$ (and $\eta_{registering}$) as wavelength increases [38]. SNSPDs based on smaller superconducting gap energy materials such as amorphous $W_xSi_{1-x}$ [36] (see section 2.3.3 and Figure 3(d)), NbSi [70] and TaN [71] appear to have better sensitivity than NbN at longer wavelengths. These low energy gap superconducting materials have lower transition temperatures, so the SNSPD operating temperature is reduced (requiring more advanced cooling techniques than those discussed in Section 2.4.1).

### 2.3.6 Resolving multi-photon events

Multi-Element SNSPDs (MESNSPDs) offer another solution to improving the single-photon count rate and achieving photon number resolution (PNR) [63, 72]. MESNSPDs consist of spatially interleaved nanowires meandering across a large active area. In this configuration each element is independently current biased and requires an individual readout [63]. This design reduces the dead time without compromising on $L_k$ or $Z_0$, or apparent cross-talk between pixels. When the pixels of the MESNSPDs are broadly illuminated photon number resolution can be achieved [72].

An alternative approach to achieving PNR in SNSPDs is by connecting multiple nanowire elements in parallel, provided the hotspot creation in one element does not affect the superconducting state of the remaining elements. This is achieved by current biasing each nanowire element independently with a single voltage source, by introducing a tailored series resistance in each nanowire element [29] (Figure 2(d)). The resistance values are carefully chosen such that the hotspot formation diverts the current into $Z_0$ instead of flowing into the remaining nanowires. When 'n' individual elements fire simultaneously, current flowing into $Z_0$ increases by a factor of 'n'. This PNR-SNSPD output results in a 'n' times taller observable voltage pulse providing PNR capability [29, 73].

Another technique to observe multiphoton events using SNSPDs is to reduce the bias current and rely on the absorption of multiple photons to form a resistive barrier across the nanowire [7, 74]. Recently new detector tomography techniques have been employed to examine this regime [75]. However, at low current bias the SNSPD is typically much less efficient, so this multiphoton detection regime is hard to exploit in practical applications.

*2.3.7 Polarization dependence in SNSPDs*

From an optical standpoint, the meander SNSPD device is essentially a subwavelength grating and is observed to possess noticeable polarization sensitivity that varies with wavelength [25]. The count rate from the detector (and hence $\eta_{sde}$) displays a maximum and a minimum value depending on whether the electric field is either parallel or perpendicular to the nanowire. This polarization dependent variation in $\eta_{absorption}$ due to the meander structure can be simulated convincingly [25]. This effect can be mitigated by using spiral and perpendicular device designs [76]. These designs eliminate the SNSPD polarisation sensitivity by accepting an intermediate efficiency.

*2.3.8 Noise mechanisms in SNSPDs: dark counts and timing jitter*

The signal-to-noise ratio, which can be achieved with an SNSPD, is governed not just by the practical efficiency ($\eta_{sde}$) and the dark count rate (DCR) but also timing jitter ($\Delta t$) [10]. The DCR of the SNSPD is a crucial factor in the device performance. Empirically, the DCR rate rises exponentially as $I_{bias}$ approaches $I_c$ (Figure 3(a))[32]. It is clear from these data (from a device with a large area meander with no cavity) at the shorter illumination wavelength ($\lambda = 830$nm) relatively high efficiency can be achieved at negligible dark count rates. The origin of the exponential behaviour of dark count rate with bias is poorly understood. Theoretical explanations centre on the possible transit of flux vortices across the current-carrying superconducting nanowire. Several mechanisms have been put forward: quantum phase-slips, single-vortex crossings and vortex-antivortex pair nucleation [77].

Experimental tests have been far from conclusive [78, 79], owing to the difficulty of differentiating these fundamental mechanisms from prosaic environmental noise owing to the bias and readout configuration. Recent studies of dark counts over a wide temperature range (0.5 – 4 K) provide valuable fresh data [34] (see Figure 3(b)). Furthermore, new theoretical studies also indicate that the device geometry (in particular, tight bends in the meander structure) may play a crucial role in triggering dark counts [80].

The very low timing jitter $\Delta t$ of SNSPDs makes these detectors very attractive for TCSPC applications. When the arrival time of the photon is known extremely well (using modern ultrafast optical sources) the main timing uncertainty arises from the jitter of the SPD. The lowest timing jitter reported is 18 ps [81], but this result has not been widely reproduced and the dimensions of the device used and photon flux were unclear. 29 ps FWHM has been demonstrated in small area (4 μm x 4.2 μm) single-and multi-pixel SNSPDs [72]. Larger area SNSPDs typically give larger timing jitter when measured with femtosecond laser sources and state-of-the-art TCSPC electronics: for example 10 μm x 10 μm meanders of the type first reported by Verevkin [27] give 68 ps FWHM [82] and 20 μm x 20 μm meander devices [32] give 60 ps FWHM (Figure 3(c)). These results indicate that a highly uniform nanowire (most easily achieved with a short wire length) will give lower timing jitter. Higher critical currents have been observed to improve jitter in uniform devices [83] allowing 40 ps FWHM to be demonstrated in 15 μm x 15 μm meander devices [84]. Recently studies using nano-optical techniques to study the local efficiency and timing properties of SNSPDs [85] indicate that different parts of a non-uniform nanowire give different hotspot resistances yielding different pulse rise times, broadening the overall timing jitter.

*2.4  Practical considerations: cooling, optical coupling and device readout*

*2.4.1  Cooling*

The most common method for cooling a SNSPD is to immerse it in liquid helium (He) [7] or to mount the device in a cryogenic probe station [28] and reach 4.2 K. Pumping on the He bath [86, 87] can reduce the temperature further (see Figure 4(a) (i)). Liquid He is expensive, hazardous and demands trained personnel for correct use. This technique is satisfactory for testing superconducting devices in a low temperature physics laboratory; however if the ultimate goal is to provide a working device for users in other scientific fields or in industrial applications, alternative cooling methods must be sought. Operating SNSPDs in a closed-cycle refrigerator [8, 88] offers a solution to this problem. The circulating fluid is high pressure, high purity He gas which is enclosed inside the refrigerator allowing continuous operation and eliminating repeated cryogenic handling. A recent version of a SNSPD system operating on this principle is shown in Figure 4 (a) (ii).

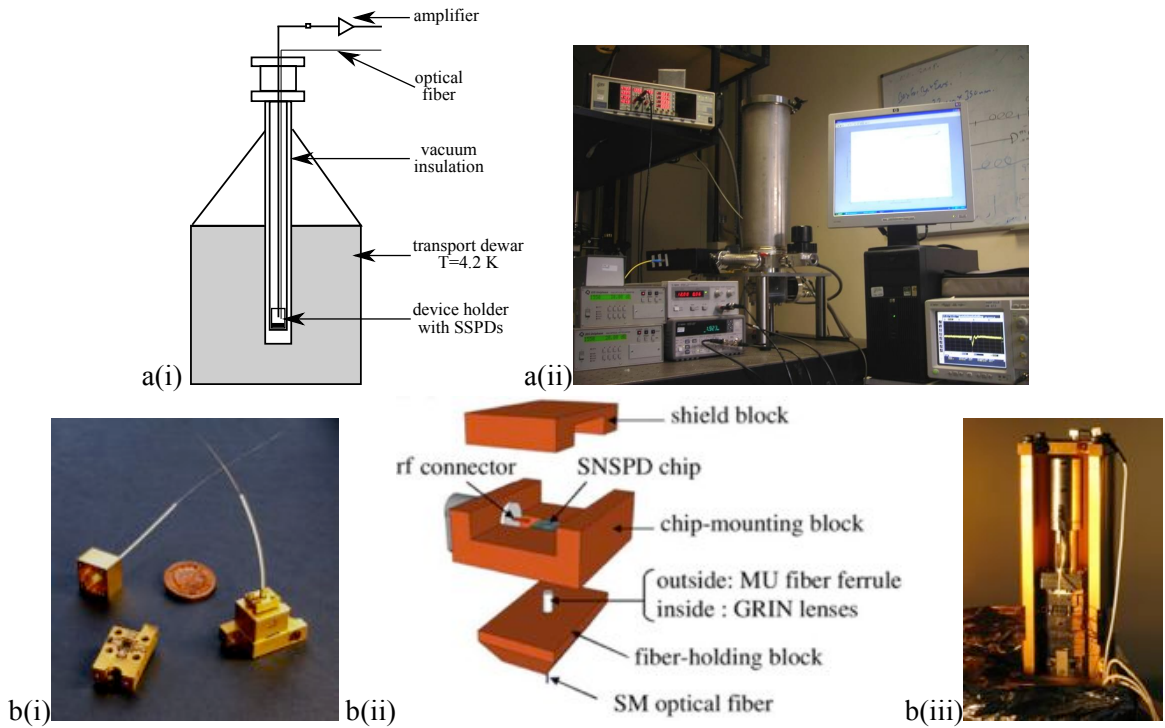

**Figure 4.** (a) Cooling techniques for SNSPDs (i) Liquid Helium dewar [87] (ii) Gifford-McMahon (GM) closed-cycle refrigerator. (b) Optical-coupling techniques (i) fibre-coupling (as used in [33, 35]) (ii) GRIN lens fibre-coupling [40] and (iii) confocal microscopy setup in a GM cryocooler (as used in [85]). Figure (a)(i) reproduced with permission from IEEE [87], (b)(ii) reproduced with permission from The Optical Society of America [40].

## 2.4.2 Optical coupling

The most obvious way to couple photons to a detector inside a cryostat is via an optical window [7]. This technique has the following disadvantages: (i) it is difficult to focus the optical spot on the 10 μm x 10 μm or 20 μm x 20 μm nanowire meander area, (ii) vibrations in the refrigerator and ambient temperature fluctuations outside the cryostat result in poor alignment stability, (iii) operating multiple SNSPDs in parallel will require separate alignment optics, this will eventually limit the number of simultaneous operating detectors and (iv) the DCR will be higher due to the increase in blackbody radiation and stray light coupling to the device. Delivering the photons through a standard single mode optical fibre [32, 86-88] (package shown in Figure 4 (b)(i)) or lensed fibre [40, 41] (package shown in Figure 4 (b) (ii)) overcomes these issues, and removes the need for an optical window in the cryostat. The fibre coupled configuration makes the operating environment more favourable for low noise SNSPD operation at low DCR, due to reduction in blackbody radiation and stray light reaching the detector. However, this technique can potentially suffer from optical misalignment, since the fibre is typically aligned to the active area of the SNSPD at room temperature and subsequently cooled down to cryogenic temperatures. New 'self-alignment' schemes [89] cleverly constrain the position of the device relative to the fibre and can be applied to SNSPDs on silicon substrates [90]. Cryogenic piezoelectric nano-positioners have been employed to provide precision *in situ* optical alignment at low temperatures. This method has helped to improve the system detection efficiency for smaller active area devices [41] (section 2.3.2), but the high cost of the nanopositioners is a barrier to widespread adoption. Cryogenic nano-positioners are also of great benefit in carrying out nano-optical studies of local device properties. Local sections within the SNSPD have been examined with sub-micrometre resolution with the help of a confocal microscopy setup (shown in Figure 4 (b) (iii)), which can be implemented either in a liquid-He cryostat [58] or a closed-cycle refrigerator [85].

*2.4.3 Readout circuits*

As discussed in section 2.3 important avenues in the advancement of SNSPD technology are the development of multi-pixel SNSPD arrays [29, 63, 72], and integration of SNSPDs with other technologies, such as on-chip optical waveguides [31, 50-52]. More efficient, low noise readout schemes are therefore of paramount importance. Progress has been made on several low-noise superconducting readout schemes. Rapid Single-Flux Quantum (RSFQ) technology [91] has been successfully integrated at low temperature with SNSPD single pixel devices and arrays [84, 92-94]. The output pulses from RSFQ circuits to room temperature require further amplification, but nevertheless impressive timing jitter has been achieved (as low as 40ps FWHM). A promising alternative to RSFQ readout is a recently demonstrated superconducting amplifier/discriminator scheme, which provides large voltage pulses [95] with low timing jitter [96].

# 3    Advanced Photon-Counting Applications with SNSPDs

*3.1    Overview*

SNSPDs are highly promising alternatives for time-correlated single photon counting (TCSPC) [9] at infrared wavelengths, offering single-photon sensitivity combined with low DCR, low $\Delta t$, short recovery times and free-running operation. Detailed comparisons have been made between SNSPDs, SPADs and other photon counting technologies in Ref [10] Table 1 and Ref [97] Table II. Over the past decade, the burgeoning field of optical quantum information science (QIS) has acted as a powerful driver for infrared single-photon detector development [98]. SNSPDs now play a significant role as an enabling technology in the QIS arena [10]. These experiments rely on the detection of coincidences between correlated photons; the detection rate of two-fold coincidences will scale as $(\eta_{sde})^2$, and the detection rate of an *n*-fold coincidence will scale as $(\eta_{sde})^n$, so as SNSPD technology improves, much more challenging experiments become feasible. However, it is important to appreciate that the scope of potential applications for SNSPDs is much wider than QIS: SNSPD technology is potentially an alternative *for any current application* of the InGaAs SPAD [12], the most widely available commercial alternative for photon-counting in the 1-1.7 μm wavelength range. In this concluding section of our Review, we survey the use of SNSPDs in several important application areas: quantum key distribution (QKD), optical quantum computing, characterization of quantum emitters, space-to-ground communications, integrated circuit testing, fibre sensing and time-of-flight ranging.

*3.2    Quantum key distribution (QKD)*

Quantum key distribution [99] (also referred to as quantum cryptography) is a method for two parties (known as Alice & Bob) to create a cryptographic key via a public communications channel. This is achieved in practice by encoding information on the phase or polarization of single photons. Any attempt by an eavesdropper (Eve) to intercept and duplicate the key will be revealed by an elevated error rate. A variety of QKD protocols and practical implementations now exist. The ideal wavelength for long distance transmission in optical fibre is 1550 nm. Therefore for long distance QKD in optical fibre, high performance single-photon detectors at λ=1550 nm are essential. A low

DCR is highly desirable, as dark counts contribute to the error rate. Low $\Delta t$ allows photon arrivals to be time-stamped very precisely. High $\eta_{sde}$ is desirable, but not essential as significant losses occur elsewhere (e.g. in the fibre link and in the receiver optics).

The first proof-of-principle demonstration using SNSPDs in QKD was carried out by the NIST and BBN groups [100]. A high bit rate, short wavelength ($\lambda$=850 nm) demonstration was then reported [101]. However, the undoubted breakthrough result was a high bit rate long distance demonstration at $\lambda$=1550 nm carried out at Stanford University [102] (Figure 5(a)). This demonstration exceeded 200 km transmission range for the first time, and record bit rates were achieved at shorter distances – a significant improvement on the best QKD results achieved at that time with InGaAs SPADs [103]. Since that study, further QKD demonstrations have been reported using SNSPDs: the maximum range has been extended to 250 km (using low loss fibre) [104], novel decoy state protocols have been demonstrated [105, 106], entanglement-based QKD has been demonstrated over long distance [107], SNSPDs have been implemented in QKD field trials in installed fibre networks [108-110] and high key transmission rates have been demonstrated [111]. Attention has turned to detector-based security loopholes in QKD - recent work [112] suggests that SNSPDs may be susceptible to 'quantum hacking' if care is not taken in the installation of the detectors in a QKD system.

*3.3  Development of optical quantum computing*

The realization of a quantum computer [113] is undoubtedly one of the grand challenges for physics in the 21$^{st}$ century, and is the focus of considerable efforts in the QIS community. The template for a scalable optical quantum computer was set out in 2001 [114]. This scheme places stringent requirements on single-photon detector technology, namely detectors with near-unity $\eta_{sde}$ and PNR capability [10]. As the performance of SNSPDs has steadily improved, SNSPDs have begun to be used in proof-of-principle QIS demonstrations, relying on the detection of multi-photon coincidences. The use of a low $\eta_{sde}$ SNSPD in tandem with a higher $\eta_{sde}$ InGaAs SPAD [115], allowed high coincidence rates to be obtained at telecom wavelengths in these challenging experiments. Key demonstrations employing SNSPDs in this way include the demonstration of the first telecom wavelength controlled-NOT gate (the basic building block of a quantum computer) [116], long distance entanglement swapping [117] and quantum storage of entanglement [118]. Another notable development in optical QIS has been the adoption of on-chip optical waveguides to replace bulk optics in quantum waveguide circuits [119]. Proof-of-principle demonstrations have been carried out using SNSPDs in conjunction with first generation quantum waveguide circuits at $\lambda$=804 nm [33] (see Figure 5(b)). The true benefit is seen however at $\lambda$=1550 nm, where much higher performance reconfigurable waveguide circuits can be employed for the first time using SNSPDs [120]. The integration of SNSPDs with waveguides (discussed in Section 2.3.2 [31, 51]) will allow these two QIS technologies to be combined on-chip.

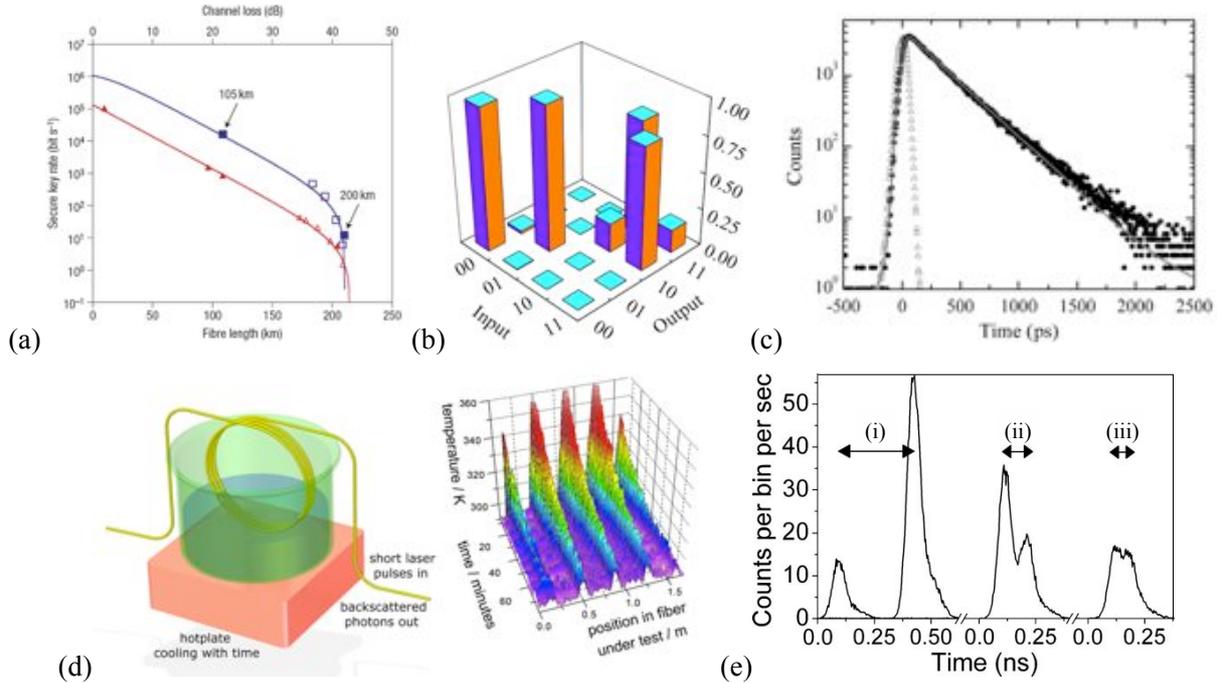

**Figure 5.** Advanced photon-counting applications with SNSPDs (a) Long distance quantum key distribution (QKD) demonstration in optical fibre using SNSPDs [102]. (b) Waveguide circuit-based CNOT gate demonstration using SNSPDs [33], (c) Photoluminescence PL measurements on a InGaAs epilayer (λ=1650 nm) using SNSPDs [121]. Black dots – PL data; triangles - SNSPD instrumental response. (d) Fibre-optic Raman temperature sensor demonstration on a fibre immersed in a water bath during cooling (after [122]). (e) Time-of-flight ranging demonstration at λ=1550 nm wavelength [123]. The target is a pair of corner cube retroreflectors separated by a distance of (i) 50 mm (ii) 15 mm and (iii) 10 mm at an overall range of 330 m in daylight. Figures reproduced with author permission: (a) NPG 2007 (b) AIP 2010 (c) Taylor & Francis 2009 (f) OSA 2008.

### 3.4 Characterization of quantum emitters

Single photon emission from atoms, quantum dots and molecules can be harnessed as a tool for production of quantum states of light for QIS and as a powerful monitoring technique in the life sciences. Considerable efforts are also underway to develop sources of correlated or entangled photons for QIS applications. SNSPDs are a powerful enabling tool for these studies at infrared wavelengths.

Individual atoms only give photon emission at wavelengths sharply defined by their energy levels [124]. III-V semiconductor quantum dots [125] act as artificial atoms, allowing the wavelength of the emission to be tuned over a wide range in the infrared (commonly λ = 850 – 1550 nm). Nitrogen vacancy defects in diamond [126] are also attractive as room temperature single-photon emitters, emitting at λ~680 nm. Single molecules are another alternative for single photon emitters, but are prey to blinking and photobleaching [127]. SNSPDs can be used in tandem with an ultrafast pump laser and TCSPC electronics to perform two key emitter characterization measurements [82, 88, 121]: firstly a measurement of the photoluminescence lifetime and secondly a measurement of the second order correlation function $g^{(2)}(0)$ in a Hanbury-Brown and Twiss configuration [125]. The low jitter and infrared sensitivity of the SNSPD allows very short lifetimes to be resolved even in wavelengths approaching 2 μm [128, 129] (see Figure 5(c)). $g^{(2)}(0)$ measurements have been successfully carried out using SNSPDs at 1.3 μm wavelength [130, 131]. SNSPDs measurements have also been reported on diamond NV centres [132] and novel solid state lasers [133]. MESNSPD designs allow new opportunities to explore higher order correlations in optical sources if multiple pixels are broadly illuminated – fascinating measurements have recently been reported [134-136].

Photon pair sources can also be combined with SPDs to create heralded single-photon sources. SNSPDs have been used in the development and characterization of various types of telecom wavelength photon pair sources: spontaneous parametric downconversion sources [115, 120], four-wave mixing sources in fibre [137, 138] and waveguide based sources [139-142]. Recently SNSPDs have also been employed in the characterization of degenerate squeezed light at telecom wavelengths as a resource for QIS [143].

*3.5  Classical space-to-ground communications*

An important area where low $\Delta t$ telecom wavelength detectors are desirable is space-to-ground communications [144]. The scenario under consideration is efficient transmission of data back to earth from a space probe with limited laser power. The SNSPD is under serious consideration for a ground-based receiver at λ=1550 nm. To date impressive data rates have been achieved (in excess of 1Gbit/s) with pulse-position-modulation and MESNSPDs [145, 146]. A full receiver (with four, 4-element SNSPD arrays) is under development at MIT Lincoln Laboratory for the NASA Lunar Laser Communication Demonstration program [147].

*3.6  Integrated circuit testing*

There is a high demand for improving practical techniques to debug and to diagnose chip failure of Complementary Metal Oxide Semiconductor (CMOS) logic devices in the semiconductor industry. In CMOS devices, when switching takes place, Field Effect Transistors (FETs) in the saturation mode develop a high field (~ $10^5$ V/cm) in the pinch off region of the conduction channel, driving electrons to high energy (> 1 eV). Photon emission takes place as the electrons lose energy. As the transistor size decreases, the gate size decreases, reducing the required bias voltage and resulting in longer wavelength emissions. Emitted photons are typically in the NIR region. The detection of these single-photons using SNSPDs enables the semiconductor industry to analyse the timing parameters of the CMOS device, and a prototype SNSPD circuit testing system has been reported [148, 149].

*3.7  Fibre temperature sensing*

With telecom wavelength single-photon sensitivity, low DCRs, low $\Delta t$ and free running operation; SNSPDs offer new opportunities in fibre sensing applications. An optical pulse passing through optical fibre is scattered to wavelengths above and below the pump wavelength via Raman scattering. SNSPDs are capable of detecting this faint backscattered signal in single-mode telecom fibre. By comparing the ratio of these Raman signals (known as the Stokes and anti-Stokes bands) it is possible to extract the temperature. With the additional timing information provided by the pump pulse and the low jitter SNSPD (coupled with TCSPC) it is possible to extract temperature as a function of length along the fibre. Initial studies [122, 150] (Figure 5(d)) show that a spatial resolution of ~1cm and a temperature uncertainty of 3 K can be achieved with integration times as short as 60 s in a test fibre ~1 m in length. With increased pump power, much shorter integration times and overall range could be achieved. Optical fibres are routinely installed in large-scale structures (buildings, pipelines) – the beauty of a fibre sensing system is that a new sensor can simply be plugged into the existing fibre infrastructure.

*3.8 Time-of-flight depth ranging*

TCSPC can be used to improve the range and performance of Light Detection and Ranging (LIDAR) systems [151]. The basic principle is as follows – a short laser pulse is transmitted and the reflected signal from a distant target is collected and routed to a single-photon detector. With a few picosecond pulsed laser, a low $\Delta t$ detector and TCPSC electronics, very good depth resolution can be achieved. SNSPDs offer an important advantage over Si SPAD detectors in this application – operation at telecom wavelengths with the SNPSDs allow solar background to be reduced and this wavelength is considered eye safe. Depth ranging studies have been carried out with SNSPDs at λ=1550 nm, achieving a depth resolution of 1 cm at 330 m range in daylight ($\eta_{sde}$ 1%, $\Delta t$ 70 ps FHWM) [123] (Figure 5(e)). In the future, scanning optics can be added to the system to achieve depth imaging [152] with similar performance. Improvements in laser power and $\eta_{sde}$ will allow accelerated imaging speeds or greater range. A very long distance (~$4 \times 10^8$ m) time-of-flight experiment will be performed as part of the Lunar Laser Communication Demonstration using SNSPDs [147] (see Section 3.5). Improving $\Delta t$ will allow better depth resolution. Moreover there is potential to use SNSPD in a wider range of atmospheric sensing applications (e.g. sensing of greenhouse gases via DIAL – Differential Absorption LIDAR [153]).

# 4 Conclusion

Over the past decade superconducting nanowire single-photon detector (SNSPD) technology has advanced apace. The basic device performance has improved dramatically, offering high $\eta_{sde}$, low DCR and low $\Delta t$ at infrared wavelengths, outperforming the best available semiconductor based single-photon detectors. Improvements in cooling technology have aided the adoption of SNSPDs in a wide range of applications. As this Review illustrates, high impact photon-counting demonstrations using SNSPDs have been carried out in diverse fields, from quantum information science to atmospheric time-of-flight ranging. There is no question that over the coming decade SNSPDs will have an important role to play in many further scientific advances and applications.

**Acknowledgements**

The authors thank colleagues who provided figures and feedback on this manuscript. The authors thank the UK Engineering and Physical Sciences Research Council (EPSRC) for support. CMN acknowledges a SU2P Entrepreneurial Fellowship and RHH acknowledges a Royal Society University Research Fellowship.